\documentclass[aps,prl,twocolumn,groupedaddress,showpacs]{revtex4}

\usepackage{graphicx}

\begin{document}

\title{Filtering Spin with Tunnel-Coupled Electron Wave Guides}

\author{M. Governale}
\author{D. Boese}
\author{U. Z\"ulicke}
\author{C. Schroll}

\affiliation{Institut f\"ur Theoretische Festk\"orperphysik,
Universit\"at Karlsruhe, D-76128 Karlsruhe, Germany}


\date{\today}

\begin{abstract}
We show how momentum-resolved tunneling between parallel electron
wave guides can be used to observe and exploit lifting of spin
degeneracy due to Rashba spin-orbit coupling. A device is proposed
that achieves spin filtering without using ferromagnets or the
Zeeman effect.
\end{abstract}


\maketitle

Spintronics is an emerging field of electronics where the
electron's spin is exploited as well as its
charge\cite{prinz:sc:98}. Operation of most spintronic devices
is based on the ability to create spin-polarized charge carriers
in nonmagnetic semiconductors. This requirement has spurred recent
interest in the investigation of possible mechanisms and
limitations of spin filtering. Simply using Zeeman splitting of
spin states is not the most practical way to achieve spin
filtering, as needed magnetic-field strengths are often large and
on-chip placement of micromagnets is required. More promising
approaches employ hybrid structures\cite{prinz:sc:90} with
metallic\cite{mj:prl:99,cam:prb:99:sh,ploog:prl:01:sh} or 
semiconducting\cite{oest:apl:99:sh,mol:nat:99:sh,ohn:nat:99:sh}
magnetic contacts. However, fabrication of these structures can
pose material-science
challenges\cite{mol:prb:00:sh,rashba:prb:00,kirc:prb:01}
and may require rather complicated chip design. Achieving spin
filtering by means of intrinsic spin-dependent effects in
semiconductors is therefore highly desirable and also very
intriguing from a fundamental-science point of view. For example,
optical excitation from spin-split hole subbands in asymmetric
quantum heterostructures can be used\cite{prettl:prl:01:sh,opt1}
to create spin-polarized currents without ferromagnets or magnetic
fields. Similarly, resonant transmission through spin-orbit-split 
quasi-bound states in semiconductor nanostructures is spin
selective and may lead to significant polarization of electron
current\cite{andra:prb:99,kis:apl:01}.

Here we propose a spin-filtering device based on the interplay of
the Rashba effect\cite{rashba,byra:jetplett:84} and wave-number
selectivity due to momentum-resolved tunneling\cite{dircoupl2}
between parallel electron wave guides. Spin-polarized currents are
created by applying voltages or small magnetic fields. Switching
between opposite spin polarizations is easily achieved.
Verification of spin filtering in the device is possible via
measurement of the differential tunneling conductance, which would
yield the first {\em direct\/} observation of Rashba-spin-split
electron dispersion curves.

\begin{figure}
\includegraphics[width=3in]{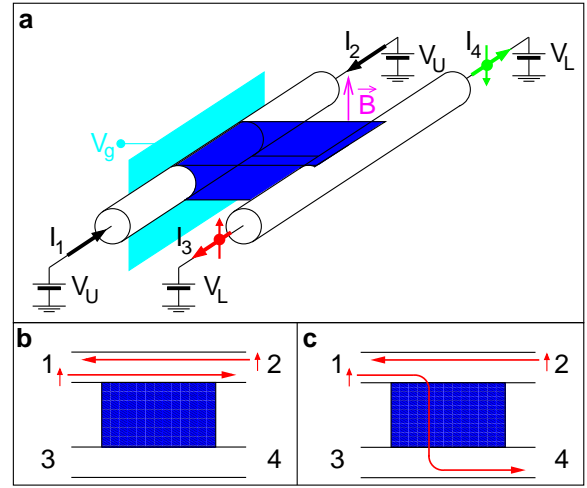}
\caption{Schematic picture of spin-filtering device and
spin-polarized currents. {\bf a,} Two parallel quantum wires,
labeled 'U' and 'L', are each connected to reservoirs having
equal chemical potential $V_{\text{U}}$ and $V_{\text{L}}$. The
wires are coupled via tunneling through an extended uniform
barrier. A gate voltage $V_{\text{g}}$ is used to control Rashba
spin-orbit coupling in the two wires. {\bf b,} Reservoir 1 (2)
injects right-moving (left-moving) electrons into the upper wire.
Shown is the situation where wave-number selectivity and
Pauli blocking prevents tunneling into the lower wire. {\bf c,}
Interplay of spin-orbit coupling and wave-number selectivity
can be used to selectively enable tunneling for right-moving
spin-up electrons. Then, a spin-polarized current is flowing from
reservoir $2$ to reservoir $4$.
\label{fig:setup}}
\end{figure}

The basic setup, shown in Fig.~\ref{fig:setup}a, consists of two
parallel one-dimensional (1D) electron wave guides (quantum wires)
that are coupled via tunneling through a clean, uniform,
nonmagnetic barrier of finite length ${\mathcal L}$. Such a system
can be realized, e.g., by quantum confinement of electrons in
semiconductor heterostructures using split-gate
techniques\cite{dircoupl2,splitdouble:sh} or cleaved-edge
overgrowth\cite{amir}. Coupling of the quantum wires via the
tunneling barrier results in a finite quantum-mechanical
probability amplitude $t_{k, k^\prime}$ for electrons to leave a
state with wave number $k$ in one wire and occupy a state with
wave number $k^\prime$ in the other wire. Translational invariance
along an extended uniform barrier implies approximate conservation
of canonical momentum in a single tunneling event. This is
seen from the explicit form of the tunneling matrix
element\cite{gov:prb:00b,uz:prb:01}, given by
\begin{equation}\label{tkp}
t_{k,k^\prime}=2 t\,\frac{\sin\left[(k^\prime-k){\mathcal L}/2
\right]}{k^\prime-k} \quad ,
\end{equation}         
whose squared modulus exhibits a delta-function-like peak for
$k=k^\prime$ with height $|t|^2 {\mathcal L}^2$ and width $2\pi/
{\mathcal L}$. The spin state of tunneling electrons remains
unchanged. Hence, tunneling between parallel quantum wires is a
highly wave-number-selective but entirely spin-insensitive
process. 

\begin{figure}
\includegraphics[width=3in]{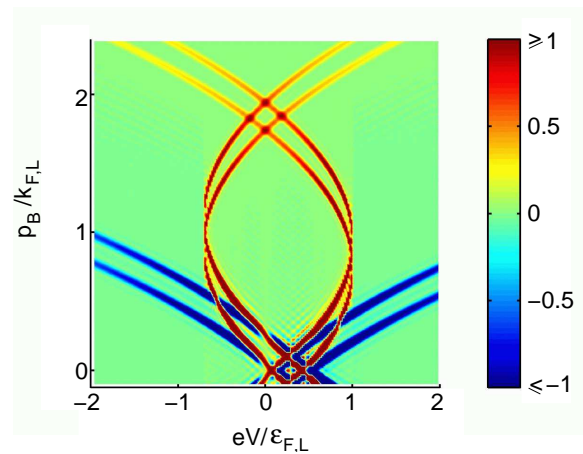}
\caption{Color plot of the differential tunneling conductance,
shown in arbitrary units, as function of voltage bias $V=
V_{\text{U}}-V_{\text{L}}$ and a magnetic field perpendicular to
the plane of the wires. The magnetic-field strength is
proportional to the parameter $p_{\text{B}}$ defined in the text.
Red and blue resonance features map out electronic dispersion
curves in the two wires. Appearance of two sets of identical
parabolic curves that are shifted in magnetic-field direction are
the unambiguous signature of Rashba spin-orbit coupling of
different strength in the two wires. For simplicity, it was
assumed in our calculation that the strength of Rashba spin-orbit
coupling is finite in the upper wire but vanishes in the lower
wire. We then use the lower wire's Fermi wave number
$k_{\text{F,L}}$ and Fermi energy $\varepsilon_{\text{F,L}}$ as
wave-number and energy units, respectively. Parameters used in
the calculation are $\Delta E_0=0.15\,\varepsilon_{\text{F,L}}$,
$k_{\text{so}}=0.1\, k_{\text{F,L}}$, ${\mathcal L}=100/
k_{\text{F,L}}$.
\label{fig:didv}}
\end{figure}

In our device, wave-number selectivity is utilized for spin
filtering by means of an intrinsic coupling of electron spin to
its momentum. Such a spin-orbit coupling originates from
structural inversion
asymmetry\cite{rashba,roess:prl:88,wink:prb:00} present in
quantum-confined systems. This effect renders the kinetic energy
of an electron with canonical momentum $\hbar k$ dependent on its
spin state\cite{byra:jetplett:84}. To illustrate the basic
physics, we consider here only the lowest 1D subband in a quantum
wire and neglect subband mixing. Then the electronic dispersion is
given by
\begin{equation}\label{rashbadisp}
E_{k\sigma} = E_0 + \frac{\hbar^2}{2 m} \left(k - \sigma
k_{\text{so}}\right)^2 - \Delta_{\text{so}} \quad .
\end{equation}
Here, the spin quantum number $\sigma$ distinguishes spin-up
($\sigma=1$) and spin-down ($\sigma=-1$) electron
eigenstates\footnote{The spin quantization axis is fixed in the
direction perpendicular to the quantum wire and the growth
direction of the 2D quantum well where it is created.}. The
effective mass of electrons is denoted by $m$, $E_0$ is the 1D
subband energy, and $\Delta_{\text{so}}=\hbar^2 k^2_{\text{so}}/(2
m)$. The strength of spin-orbit coupling can be expressed in terms
of a characteristic wave number, denoted here by $k_{\text{so}}$.
Experimental
efforts\cite{nitta:prl:97,schaep:jap:98:sh,grund:prl:00} aimed at
the realization of an early proposal\cite{spinfet} for a
spin-controlled field-effect transistor established tunability of
$k_{\text{so}}$ by external gate voltages. In our device (see
Fig.~\ref{fig:setup}a), the voltage $V_{\text{g}}$ is used to
achieve different strengths of Rashba spin-orbit coupling in the
two wires. In that situation, tunneling transport across the
barrier provides a direct measurement of the Rashba effect. The
differential tunneling conductance, calculated using standard
perturbation theory within the tunneling-Hamiltonian
formalism\cite{mahan,uz:prb:01}, is shown in Fig.~\ref{fig:didv}.
It provides a direct image of the spin-resolved parabolic
dispersion curves given by Eq.~(\ref{rashbadisp}). Monitoring the
differential tunneling conductance would be the most immediate
possibility to observe and study spectral consequences of the
Rashba effect in 1D. So far, experimental
studies\cite{nitta:prl:97,schaep:jap:98:sh,grund:prl:00} have
focused on extracting the value of Rashba-induced zero-field spin
splitting in 2D systems from the analysis of beating patterns in
the Shubnikov--de~Haas oscillations.

When a picture like Fig.~\ref{fig:didv} is obtained for the
differential tunneling conductance, the double-wire system can be
used for spin filtering. To simplify the explanation of its basic
operational modes as a spin polarizer and a spin splitter, we
consider here the special case where Rashba spin-orbit coupling
is finite in the upper wire (labeled U) but vanishes in the lower
wire (labeled L)\footnote{A finite difference $\Delta
k_{\text{so}}=k_{\text{so}}^{\text{(U)}}-
k_{\text{so}}^{\text{(L)}}$ and sufficient wave-number selectivity
ensure functionality of the device. Subsequent formulae remain
valid when $k_{\text{so}}$ is replaced by
$\Delta k_{\text{so}}$.}.

\begin{figure}
\includegraphics[width=3in]{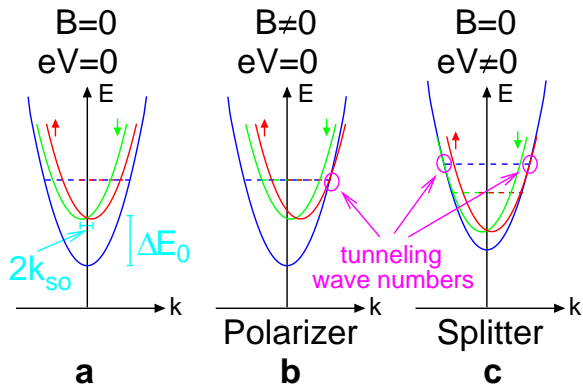}
\caption{Illustration of device operation as spin polarizer or
spin splitter {\bf a,} Due to the Rashba effect, dispersion curves
for spin-up and spin-down electrons in the upper wire are shifted
horizontally by $2 k_{\mathrm{so}}$. In the lower wire, where
spin-orbit coupling is assumed to be absent, energy dispersions
are spin-degenerate. {\bf b,} Tuning wave-number selectivity by a
magnetic field $B$, tunneling is selectively enabled for
right-moving electrons with spin up. {\bf c,} At a certain value
of voltage $V$, tunneling becomes possible for left-moving
spin-down electrons and right-moving spin-up electrons. Note that
parabolicity of electron bands is not essential to achieve
coincidences and, hence, spin-polarized currents.\label{fig:disp}}
\end{figure}

\begin{figure}
\includegraphics[width=3in]{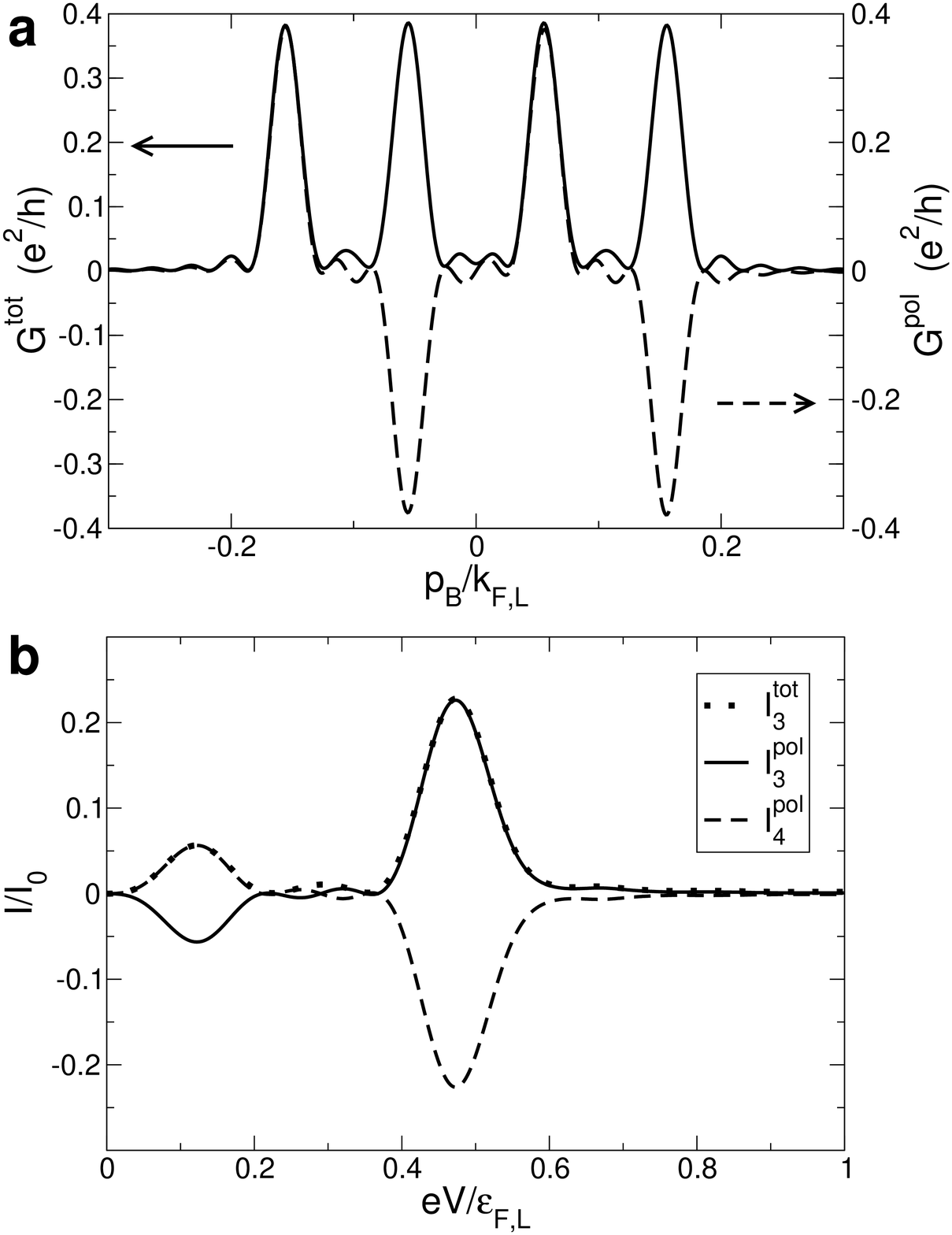}
\caption{Tunneling transport calculated exactly using scattering
theory (a) and perturbatively (b) as explained in
Ref.~\onlinecite{uz:prb:01} {\bf a,} Total linear tunneling
conductance through the barrier $G^{\text{tot}}=G_{\uparrow}+
G_{\downarrow}$ (left axis) and spin-polarized conductance
$G^{\text{pol}}=G_{\uparrow}-G_{\downarrow}$ (right axis) vs.\
magnetic field $B$. Resonances with definite spin polarization
occur at values of the magnetic field determined by
Eq.~(\ref{resonances}). Data shown are for $\Delta E_0= 0.1\,
\varepsilon_{\text{F,L}}$, $k_{\text{so}}=0.05\, k_{\text{F,L}}$,
${\mathcal L}=200 / k_{\text{F,L}}$, $|t| =
\varepsilon_{\text{F,L}}\cdot\pi/
1000$. {\bf b,} Currents in all the leads are spin-polarized when
operating the device in spin-splitting mode (Fig.~3c). Total and
spin currents entering lead $j$ are denoted by $I_j^{\text{tot}}=
I_j^{\uparrow}+I_j^{\downarrow}$ and $I_j^{\text{pol}}=\left[
I_j^{\uparrow}-I_j^{\downarrow}\right]{\mathrm{sgn}}
(I_j^{\text{tot}})$, respectively. In the spin-splitter mode, we
have $I_4^{\text{tot}}=I_3^{\text{tot}}=-I_2^{\text{tot}}=-
I_1^{\text{tot}}$. We show $I_3^{\text{tot}}$ vs.\ bias voltage
$V=V_{\text{U}}-V_{\text{L}}$ (dotted curve). It is appreciable
only in finite ranges of voltage where states near points of
coincidence for the wires' dispersion curves are occupied in one
wire but empty in the other one. Within these voltage intervals,
total current in each lead is practically 100\% polarized. This is
seen from comparing $I_3^{\text{tot}}$ with $I_3^{\text{pol}}$
(solid curve) and $I_4^{\text{pol}}$ (dashed curve). Data shown
are calculated for $\Delta E_0 = 0.15\,\varepsilon_{\text{F,L}}$,
$k_{\mathrm{so}}=0.1\, k_{\text{F,L}}$, ${\mathcal L}=100/
k_{\text{F,L}}$, $|t| = \varepsilon_{\text{F,L}}\cdot\pi/1000$,
$B=0$. Current unit is $I_0=2 e \pi |t|^2 {\mathcal L}^2 /(\hbar^3
v_{\text{F,L}}
v_{\text{F,U}})$, where $v_{\text{F,U}}$ ($v_{\text{F,L}})$ is the
Fermi velocity in the U (L) wire.
\label{fig:currents}}
\end{figure}

Taking into account a finite subband-energy
difference $\Delta E_0=E_0^{\text{(U)}}-E_0^{\text{(L)}}$,
electronic dispersion relations look like shown in
Fig.~\ref{fig:disp}a. Only states with energies below
$E_0^{\text{(U)}}+\varepsilon_{\text{F,U}}$ in the upper wire
(below $E_0^{\text{(L)}}+\varepsilon_{\text{F,L}}$ in the lower
wire) are occupied by electrons. Due to wave-number selectivity,
tunneling can only occur for electron states with wave number
close to a point where the dispersion curves of the two wires
cross\cite{uz:prb:01}. In the case depicted in
Fig.~\ref{fig:disp}a, no such crossings occur for states that are
occupied in either one of the two wires, and no tunneling current
can flow. All electrons injected into the upper wire from
reservoirs 1 and 2 remain in the upper wire
(Fig.~\ref{fig:setup}b). This situation can change when a magnetic
field $B$ is applied perpendicular to the plane of the two wires.
It leads to a relative shift\cite{uz:prb:01} of the two wires'
dispersion curves in wave-number direction by $p_{\text{B}}=-e B d
/\hbar$, where $-e$ is the electron charge, and $d$ the wire
separation. For particular values of magnetic field, namely
whenever the condition
\begin{equation}\label{resonances}
p_{\text{B}}=\frac{\pi}{2} \left(\gamma n_{\text{U}} -
\gamma^\prime n_{\text{L}} \right) - \sigma k_{\text{so}}
\end{equation}
is satisfied with $\gamma,\gamma^\prime=\pm 1$ and
$n_{\text{U(L)}}$ denoting electron density in the upper (lower)
wire, tunneling is enabled for electrons in spin state $\sigma$.
Simultaneous tunneling of electrons with opposite spin will be
prohibited for high enough wave-number selectivity, i.e., if
$\pi/{\mathcal L}\ll k_{\text{so}}$. The case corresponding to
$\gamma=\gamma^\prime=+1$ is depicted in Fig.~\ref{fig:disp}b. In
that situation, a spin-polarized current of electrons from
reservoir 1 reaches reservoir 4. As shown in
Fig.~\ref{fig:setup}c, it is
compensated globally by the current of spin-up electrons from
reservoir 2 that can only reach reservoir 1. Using the standard
scattering-theory formalism\cite{butt:ibm:88} for calculating
electron transport, we have obtained the linear conductance for
tunneling across the barrier. Our results, given in
Fig.~\ref{fig:currents}a, show the four resonances determined by
solutions of Eq.~(\ref{resonances}) which exhibit almost perfect
spin polarization of the tunneling current. We emphasize that the
applied magnetic field, used here simply to tune wave-number
selectivity, is typically much smaller than the field required to
achieve spin-filtering from Zeeman splitting. Tuning between the
four resonances allows the double-quantum-wire system to be used
as a switchable spin-polarizer. Analyzing the resonance condition
given by Eq.~(\ref{resonances}) from a fundamental point of view,
we realize that this is possible because Rashba spin-orbit
coupling enters the single--particle Hamiltonian like a
spin-dependent vector potential.

Intriguingly, spin-polarized currents can be created in our device
without applying any magnetic field at all. Instead, a finite bias
voltage $V=V_{\text{U}}-V_{\text{L}}$ can be used to induce a
relative shift $-e V$ of the two wires' dispersion curves in
energy direction\footnote{In general, charging and
exchange-correlation effects in the wires renormalize the
voltage-induced shift of their dispersion curves. This can be
taken into account straightforwardly and does not alter
functionality of our device as a spin-splitter.}. This way,
coincidences are created for electron states with spin $\sigma$
and wave number $k$ satisfying
\begin{equation}\label{voltcond}
e V = \Delta E_0 - \frac{\hbar^2}{m}\, k_{\text{so}}\,\sigma\, k
\quad .
\end{equation}
If such a state is occupied by an electron in one wire and empty
in the other one, it will contribute to the tunneling current. A
situation where current flow is made possible by the applied
voltage is depicted in Fig.~\ref{fig:disp}c. Tunneling is
simultaneously enabled for electrons having opposite spin and wave
number, which will end up in opposite leads of the wire they have
tunneled into. For large enough wave-number selectivity, tunneling
for a particular spin species turns out to occur, if at all, only
for wave numbers of one sign. As a result, currents flowing in the
leads of the double-wire device are fully spin-polarized. This is
seen in Fig.~\ref{fig:currents}b for a particular set of
parameters. Currents in leads 1 and 3 have the same spin
polarization, which is opposite to that in leads 2 and 4.
Selection of spin-up or spin-down polarization for currents in
leads 1 and 3 (2 and 4) is possible simply by adjusting the
voltage. While no global spin imbalance is created, wave-number
selectivity leads to a redistribution of spin-polarized currents
between the four leads. Thereby, spin filtering is possible
without any magnetic or exchange fields. Note that wave-number
selectivity provides the most direct way to utilize the Rashba
effect for spin filtering. Hence, besides opening up an
interesting alternative to previously suggested energy-selective
mechanisms\cite{kis:apl:01}, our device offers certain advantages
that may be important for application in spintronics\footnote{The
most important advantage is that spin filtering occurs in our
setup when the transmission into output channels is maximal. This
is different in the T-junction device of \citet{kis:apl:01} where
resonance with quasi-zerodimensional bound states is essential to
achieve appreciable spin filtering but, at the same time, results
in almost complete backscattering.}.

Three mechanisms limit functionality of a real double-wire system
as a spin-filtering device. First, wave-number selectivity is
reduced by disorder. However, the successful measurement of 1D
dispersion curves for parallel quantum wires in GaAlAs
heterostructures\cite{amir} demonstrates the possibility to
achieve sufficient wave-number selectivity using present-day
technology. While this particular system cannot be used for spin
filtering due to its negligible Rashba effect, similar structures
could be created in more suitable materials. Second, quantum wires
are really multi-channel 1D wave guides for electrons. Spin-orbit
coupling induces mixing between these channels (subbands), which
was neglected in our discussion so far. Detailed
analysis\cite{spinfet} shows that subband mixing can be safely
neglected as long as the energy difference between consecutive
quasi-1D subbands is much larger than $\Delta_{\text{so}}$. In
present-day samples, this requirement is easily met for realistic
values\cite{nitta:prl:97,schaep:jap:98:sh,grund:prl:00} of
$k_{\text{so}}$. Finally, our conclusions apply at temperatures
$T$ low enough such that smearing of the Fermi function does not
substantially decrease wave-number selectivity. The relevant
criterion $k_{\text{B}} T\lesssim\hbar^2 k_{\text{so}}
k_{\text{F,L}}/m$ translates, for typical sample parameters, into
$T\lesssim 10$~K. This temperature range is routinely accessible
in semiconductor-research laboratories where our basic design for
a spin-filtering device could be demonstrated.

Experimental tests for the successful operation of our device as
spin polarizer or spin splitter could be based on the usual
spin-detection mechanisms, applied to currents leaving or
entering any of the four leads. For example, ferromagnetic
contacts can serve as spin analyzers, or spin polarization of
charge carriers could be measured
optically\cite{ploog:prl:01:sh,oest:apl:99:sh,mol:nat:99:sh,ohn:nat:99:sh}.
Results of such experiments will be determined not only by the
efficiency of spin filtering in the double-wire system but also by
spin-equilibration processes in the leads. An equally significant
test is provided by measurement of the differential tunneling
conductance as a function of magnetic field and transport
voltage. Clearly resolved Rashba-split dispersion curves, as
shown in Fig.~\ref{fig:didv}, prove sufficient wave-number
selectivity for addressing different spin states and constitute
therefore the decisive experimental demonstration of spin
filtering in our device.

This work was supported by Deutsche Forschungsgemeinschaft within
the Center for Functional Nanostructures and Graduiertenkolleg
"Kollektive Ph\"anomene im Festk\"orper". We thank E.~Rashba,
G.~Sch\"on, and B.~van~Wees for useful discussions.


\begin{thebibliography}{30}
\expandafter\ifx\csname natexlab\endcsname\relax\def\natexlab#1{#1}\fi
\expandafter\ifx\csname bibnamefont\endcsname\relax
  \def\bibnamefont#1{#1}\fi
\expandafter\ifx\csname bibfnamefont\endcsname\relax
  \def\bibfnamefont#1{#1}\fi
\expandafter\ifx\csname citenamefont\endcsname\relax
  \def\citenamefont#1{#1}\fi
\expandafter\ifx\csname url\endcsname\relax
  \def\url#1{\texttt{#1}}\fi
\expandafter\ifx\csname urlprefix\endcsname\relax\def\urlprefix{URL }\fi
\providecommand{\bibinfo}[2]{#2}
\providecommand{\eprint}[2][]{\url{#2}}

\bibitem[{\citenamefont{Prinz}(1998)}]{prinz:sc:98}
\bibinfo{author}{\bibfnamefont{G.~A.} \bibnamefont{Prinz}},
  \bibinfo{journal}{Science} \textbf{\bibinfo{volume}{282}},
  \bibinfo{pages}{1660} (\bibinfo{year}{1998}).

\bibitem[{\citenamefont{Prinz}(1990)}]{prinz:sc:90}
\bibinfo{author}{\bibfnamefont{G.~A.} \bibnamefont{Prinz}},
  \bibinfo{journal}{Science} \textbf{\bibinfo{volume}{250}},
  \bibinfo{pages}{1092} (\bibinfo{year}{1990}).

\bibitem[{\citenamefont{Hammar et~al.}(1999)\citenamefont{Hammar, Bennett,
  Yang, and Johnson}}]{mj:prl:99}
\bibinfo{author}{\bibfnamefont{P.~R.} \bibnamefont{Hammar}},
  \bibinfo{author}{\bibfnamefont{B.~R.} \bibnamefont{Bennett}},
  \bibinfo{author}{\bibfnamefont{M.~J.} \bibnamefont{Yang}}, \bibnamefont{and}
  \bibinfo{author}{\bibfnamefont{M.}~\bibnamefont{Johnson}},
  \bibinfo{journal}{Phys. Rev. Lett.} \textbf{\bibinfo{volume}{83}},
  \bibinfo{pages}{203} (\bibinfo{year}{1999}).

\bibitem[{\citenamefont{Gardelis et~al.}(1999)}]{cam:prb:99:sh}
\bibinfo{author}{\bibfnamefont{S.}~\bibnamefont{Gardelis}}
  \bibnamefont{et~al.}, \bibinfo{journal}{Phys. Rev. B}
  \textbf{\bibinfo{volume}{60}}, \bibinfo{pages}{7764} (\bibinfo{year}{1999}).

\bibitem[{\citenamefont{Zhu et~al.}(2001)}]{ploog:prl:01:sh}
\bibinfo{author}{\bibfnamefont{H.~J.} \bibnamefont{Zhu}} \bibnamefont{et~al.},
  \bibinfo{journal}{Phys. Rev. Lett.} \textbf{\bibinfo{volume}{87}},
  \bibinfo{pages}{016601} (\bibinfo{year}{2001}).

\bibitem[{\citenamefont{Oestreich et~al.}(1999)}]{oest:apl:99:sh}
\bibinfo{author}{\bibfnamefont{M.}~\bibnamefont{Oestreich}}
  \bibnamefont{et~al.}, \bibinfo{journal}{Appl. Phys. Lett.}
  \textbf{\bibinfo{volume}{74}}, \bibinfo{pages}{1251} (\bibinfo{year}{1999}).

\bibitem[{\citenamefont{Fielderling et~al.}(1999)}]{mol:nat:99:sh}
\bibinfo{author}{\bibfnamefont{R.}~\bibnamefont{Fielderling}}
  \bibnamefont{et~al.}, \bibinfo{journal}{Nature (London)}
  \textbf{\bibinfo{volume}{402}}, \bibinfo{pages}{787} (\bibinfo{year}{1999}).

\bibitem[{\citenamefont{Ohno et~al.}(1999)}]{ohn:nat:99:sh}
\bibinfo{author}{\bibfnamefont{Y.}~\bibnamefont{Ohno}} \bibnamefont{et~al.},
  \bibinfo{journal}{Nature (London)} \textbf{\bibinfo{volume}{402}},
  \bibinfo{pages}{790} (\bibinfo{year}{1999}).

\bibitem[{\citenamefont{Schmidt et~al.}(2000)}]{mol:prb:00:sh}
\bibinfo{author}{\bibfnamefont{G.}~\bibnamefont{Schmidt}} \bibnamefont{et~al.},
  \bibinfo{journal}{Phys. Rev. B} \textbf{\bibinfo{volume}{62}},
  \bibinfo{pages}{R4790} (\bibinfo{year}{2000}).

\bibitem[{\citenamefont{Rashba}(2000)}]{rashba:prb:00}
\bibinfo{author}{\bibfnamefont{E.~I.} \bibnamefont{Rashba}},
  \bibinfo{journal}{Phys. Rev. B} \textbf{\bibinfo{volume}{62}},
  \bibinfo{pages}{R16267} (\bibinfo{year}{2000}).

\bibitem[{\citenamefont{Kirczenow}(2001)}]{kirc:prb:01}
\bibinfo{author}{\bibfnamefont{G.}~\bibnamefont{Kirczenow}},
  \bibinfo{journal}{Phys. Rev. B} \textbf{\bibinfo{volume}{63}},
  \bibinfo{pages}{054422} (\bibinfo{year}{2001}).

\bibitem[{\citenamefont{Ganichev et~al.}(2001)}]{prettl:prl:01:sh}
\bibinfo{author}{\bibfnamefont{S.~D.} \bibnamefont{Ganichev}}
  \bibnamefont{et~al.}, \bibinfo{journal}{Phys. Rev. Lett.}
  \textbf{\bibinfo{volume}{86}}, \bibinfo{pages}{4358} (\bibinfo{year}{2001}).

\bibitem[{\citenamefont{Mal'shukov and Chao}()}]{opt1}
\bibinfo{author}{\bibfnamefont{A.~G.} \bibnamefont{Mal'shukov}}
  \bibnamefont{and} \bibinfo{author}{\bibfnamefont{K.~A.} \bibnamefont{Chao}},
  \bibinfo{note}{cond-mat/0108326}.

\bibitem[{\citenamefont{{de Andrada e Silva} and {La
  Rocca}}(1999)}]{andra:prb:99}
\bibinfo{author}{\bibfnamefont{E.~A.} \bibnamefont{{de Andrada e Silva}}}
  \bibnamefont{and} \bibinfo{author}{\bibfnamefont{G.~C.} \bibnamefont{{La
  Rocca}}}, \bibinfo{journal}{Phys. Rev. B} \textbf{\bibinfo{volume}{59}},
  \bibinfo{pages}{R15583} (\bibinfo{year}{1999}).

\bibitem[{\citenamefont{Kiselev and Kim}(2001)}]{kis:apl:01}
\bibinfo{author}{\bibfnamefont{A.~A.} \bibnamefont{Kiselev}} \bibnamefont{and}
  \bibinfo{author}{\bibfnamefont{K.~W.} \bibnamefont{Kim}},
  \bibinfo{journal}{Appl. Phys. Lett.} \textbf{\bibinfo{volume}{78}},
  \bibinfo{pages}{775} (\bibinfo{year}{2001}).

\bibitem[{\citenamefont{Rashba}(1960)}]{rashba}
\bibinfo{author}{\bibfnamefont{E.~I.} \bibnamefont{Rashba}},
  \bibinfo{journal}{Fiz. Tverd. Tela (Leningrad)} \textbf{\bibinfo{volume}{2}},
  \bibinfo{pages}{1224} (\bibinfo{year}{1960}), \bibinfo{note}{[Sov. Phys.
  Solid State {\bf 2}, 1109 (1960)]}.

\bibitem[{\citenamefont{Bychkov and Rashba}(1984)}]{byra:jetplett:84}
\bibinfo{author}{\bibfnamefont{Y.~A.} \bibnamefont{Bychkov}} \bibnamefont{and}
  \bibinfo{author}{\bibfnamefont{E.~I.} \bibnamefont{Rashba}},
  \bibinfo{journal}{Pis'ma Zh. \'Eksp. Teor. Fiz.}
  \textbf{\bibinfo{volume}{39}}, \bibinfo{pages}{66} (\bibinfo{year}{1984}),
  \bibinfo{note}{[JETP Lett. {\bf 39}, 78 (1984)]}.

\bibitem[{\citenamefont{Eugster et~al.}(1994)\citenamefont{Eugster, del Alamo,
  Rooks, and Melloch}}]{dircoupl2}
\bibinfo{author}{\bibfnamefont{C.~C.} \bibnamefont{Eugster}},
  \bibinfo{author}{\bibfnamefont{J.~A.} \bibnamefont{del Alamo}},
  \bibinfo{author}{\bibfnamefont{M.~J.} \bibnamefont{Rooks}}, \bibnamefont{and}
  \bibinfo{author}{\bibfnamefont{M.~R.} \bibnamefont{Melloch}},
  \bibinfo{journal}{Appl. Phys. Lett.} \textbf{\bibinfo{volume}{64}},
  \bibinfo{pages}{3157} (\bibinfo{year}{1994}).

\bibitem[{\citenamefont{Thomas et~al.}(1999)}]{splitdouble:sh}
\bibinfo{author}{\bibfnamefont{K.~J.} \bibnamefont{Thomas}}
  \bibnamefont{et~al.}, \bibinfo{journal}{Phys. Rev. B}
  \textbf{\bibinfo{volume}{59}}, \bibinfo{pages}{12252} (\bibinfo{year}{1999}).

\bibitem[{\citenamefont{Auslaender et~al.}(2002)}]{amir}
\bibinfo{author}{\bibfnamefont{O.~M.} \bibnamefont{Auslaender}}
  \bibnamefont{et~al.}, \bibinfo{journal}{Science}
  \textbf{\bibinfo{volume}{295}}, \bibinfo{pages}{825} (\bibinfo{year}{2002}).

\bibitem[{\citenamefont{Governale et~al.}(2000)\citenamefont{Governale,
  Grifoni, and Sch\"on}}]{gov:prb:00b}
\bibinfo{author}{\bibfnamefont{M.}~\bibnamefont{Governale}},
  \bibinfo{author}{\bibfnamefont{M.}~\bibnamefont{Grifoni}}, \bibnamefont{and}
  \bibinfo{author}{\bibfnamefont{G.}~\bibnamefont{Sch\"on}},
  \bibinfo{journal}{Phys. Rev. B} \textbf{\bibinfo{volume}{62}},
  \bibinfo{pages}{15996} (\bibinfo{year}{2000}).

\bibitem[{\citenamefont{Boese et~al.}(2001)\citenamefont{Boese, Governale,
  Rosch, and Z\"ulicke}}]{uz:prb:01}
\bibinfo{author}{\bibfnamefont{D.}~\bibnamefont{Boese}},
  \bibinfo{author}{\bibfnamefont{M.}~\bibnamefont{Governale}},
  \bibinfo{author}{\bibfnamefont{A.}~\bibnamefont{Rosch}}, \bibnamefont{and}
  \bibinfo{author}{\bibfnamefont{U.}~\bibnamefont{Z\"ulicke}},
  \bibinfo{journal}{Phys. Rev. B} \textbf{\bibinfo{volume}{64}},
  \bibinfo{pages}{085315} (\bibinfo{year}{2001}).

\bibitem[{\citenamefont{Lommer et~al.}(1988)\citenamefont{Lommer, Malcher, and
  R\"ossler}}]{roess:prl:88}
\bibinfo{author}{\bibfnamefont{G.}~\bibnamefont{Lommer}},
  \bibinfo{author}{\bibfnamefont{F.}~\bibnamefont{Malcher}}, \bibnamefont{and}
  \bibinfo{author}{\bibfnamefont{U.}~\bibnamefont{R\"ossler}},
  \bibinfo{journal}{Phys. Rev. Lett.} \textbf{\bibinfo{volume}{60}},
  \bibinfo{pages}{728} (\bibinfo{year}{1988}).

\bibitem[{\citenamefont{Winkler}(2000)}]{wink:prb:00}
\bibinfo{author}{\bibfnamefont{R.}~\bibnamefont{Winkler}},
  \bibinfo{journal}{Phys. Rev. B} \textbf{\bibinfo{volume}{62}},
  \bibinfo{pages}{4245} (\bibinfo{year}{2000}).

\bibitem[{\citenamefont{Nitta et~al.}(1997)\citenamefont{Nitta, Akazaki,
  Takayanagi, and Enoki}}]{nitta:prl:97}
\bibinfo{author}{\bibfnamefont{J.}~\bibnamefont{Nitta}},
  \bibinfo{author}{\bibfnamefont{T.}~\bibnamefont{Akazaki}},
  \bibinfo{author}{\bibfnamefont{H.}~\bibnamefont{Takayanagi}},
  \bibnamefont{and} \bibinfo{author}{\bibfnamefont{T.}~\bibnamefont{Enoki}},
  \bibinfo{journal}{Phys. Rev. Lett.} \textbf{\bibinfo{volume}{78}},
  \bibinfo{pages}{1335} (\bibinfo{year}{1997}).

\bibitem[{\citenamefont{Sch\"apers et~al.}(1998)}]{schaep:jap:98:sh}
\bibinfo{author}{\bibfnamefont{T.}~\bibnamefont{Sch\"apers}}
  \bibnamefont{et~al.}, \bibinfo{journal}{J. Appl. Phys.}
  \textbf{\bibinfo{volume}{83}}, \bibinfo{pages}{4324} (\bibinfo{year}{1998}).

\bibitem[{\citenamefont{Grundler}(2000)}]{grund:prl:00}
\bibinfo{author}{\bibfnamefont{D.}~\bibnamefont{Grundler}},
  \bibinfo{journal}{Phys. Rev. Lett.} \textbf{\bibinfo{volume}{84}},
  \bibinfo{pages}{6074} (\bibinfo{year}{2000}).

\bibitem[{\citenamefont{Datta and Das}(1990)}]{spinfet}
\bibinfo{author}{\bibfnamefont{S.}~\bibnamefont{Datta}} \bibnamefont{and}
  \bibinfo{author}{\bibfnamefont{B.}~\bibnamefont{Das}},
  \bibinfo{journal}{Appl. Phys. Lett.} \textbf{\bibinfo{volume}{56}},
  \bibinfo{pages}{665} (\bibinfo{year}{1990}).

\bibitem[{\citenamefont{Mahan}(1990)}]{mahan}
\bibinfo{author}{\bibfnamefont{G.~D.} \bibnamefont{Mahan}},
  \emph{\bibinfo{title}{Many-Particle Physics}} (\bibinfo{publisher}{Plenum
  Press}, \bibinfo{address}{New York}, \bibinfo{year}{1990}).

\bibitem[{\citenamefont{B\"uttiker}(1988)}]{butt:ibm:88}
\bibinfo{author}{\bibfnamefont{M.}~\bibnamefont{B\"uttiker}},
  \bibinfo{journal}{IBM J. Res. Dev.} \textbf{\bibinfo{volume}{32}},
  \bibinfo{pages}{317} (\bibinfo{year}{1988}).

\end{thebibliography}

\end{document}